\documentclass[prl,aps,twocolumn,groupedaddress,showpacs]{revtex4}

\usepackage{graphicx}

\begin{document}

\title{Observation of Phase Separation in a Strongly-Interacting Imbalanced Fermi Gas}

\author{Y. Shin}
\email{yishin@mit.edu}
\author{M.~W. Zwierlein}
\author{C.~H. Schunck}
\author{A. Schirotzek}
\author{W. Ketterle}

\affiliation{Department of Physics, MIT-Harvard Center for
Ultracold Atoms, and Research Laboratory of Electronics,
Massachusetts Institute of Technology, Cambridge, Massachusetts,
02139}

\date{\today}

\begin{abstract}
We have observed phase separation between the superfluid and the
normal component in a strongly interacting Fermi gas with
imbalanced spin populations. The in situ distribution of the
density difference between two trapped spin components is
obtained using phase-contrast imaging and 3D image
reconstruction. A shell structure is clearly identified where the
superfluid region of equal densities is surrounded by a normal
gas of unequal densities. The phase transition induces a dramatic
change in the density profiles as excess fermions start being
expelled from the superfluid.
\end{abstract}


\maketitle

Cooper pairing is the underlying mechanism for the
Bardeen-Cooper-Schrieffer (BCS) superfluid state of an equal
mixture of two fermionic gases. An interesting situation arises
when the two components have unequal populations. Does the
imbalance quench superfluidity, does it lead to phase separation
between a balanced and imbalanced region, or does it give rise to
new forms of superfluidity? A search for exotic superfluid states
is promising in imbalanced mixtures, since the imbalance
destabilizes BCS-type $s$-wave pairing which is usually the
strongest pairing mechanism~\cite{FF64,LO65,Sar63,LW03}.
Recently, this problem has been experimentally addressed in
ultracold atomic Fermi clouds with controlled population
imbalances~\cite{ZSS06a,PLK06,ZSS06b}. Superfluidity was observed
in a strongly interacting regime with a broad range of imbalances
and the Pauli or Clogston limit of superfluidity~\cite{CLO62} was
characterized~\cite{ZSS06a,ZSS06b}.

The phase separation scenario suggests that unpaired fermions are
spatially separated from a BCS-superfluid of equal densities due
to the pairing gap in the superfluid
region~\cite{BCR03,CR05,SR06}. In our previous
experiments~\cite{ZSS06a,ZSS06b}, we observed a strong central
depletion in the difference profiles of expanding clouds
indicating that excess atoms are expelled from the superfluid
region. Ref.~\cite{PLK06} reports depletion of excess atoms at
the trap center. It has been clarified~\cite{ZK06,PLK06b} that
none of these experiments answered the question: whether the
densities of the two spin components are \emph{equal} in the
\emph{superfluid} region and whether phase separation or rather
distortions of the cloud due to interactions have occurred.

Here we report the direct observation of phase separation between
the superfluid and the normal region in a strongly interacting
Fermi gas with imbalanced spin populations. The density
difference between the two spin components is directly measured
\emph{in situ} using a special phase-contrast imaging technique
and 3D image reconstruction. We clearly identify a shell
structure in an imbalanced Fermi gas where the superfluid region
of equal densities is surrounded by a normal gas of unequal
densities. This phase separation is observed throughout the
strongly interacting regime near a Feshbach resonance.
Furthermore, we characterize the normal-to-superfluid phase
transition of an imbalanced Fermi mixture using in situ
phase-contrast imaging. The onset of superfluidity induces a
dramatic change in the density profiles as excess fermions are
expelled from the superfluid.

\begin{figure}
\begin{center}
\includegraphics[width=2.7in]{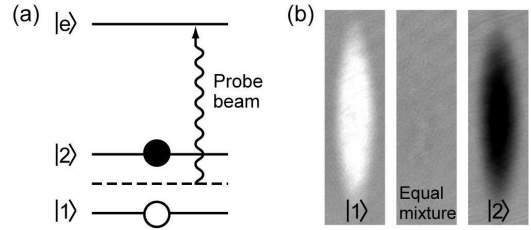}
\caption{Phase-contrast imaging of the density difference of two
spin states. (a) The probe beam is tuned to the red for the
$|1\rangle\rightarrow|e\rangle$ transition and to the blue for the
$|2\rangle\rightarrow|e\rangle$ transition. The resulting optical
signal in the phase-contrast image is proportional to the density
difference $n_d\equiv n_1 - n_2$, where $n_1$ and $n_2$ are the
densities of the states $|1\rangle$ and $|2\rangle$,
respectively. (b) Phase-contrast images of trapped atomic clouds
in state $|1\rangle$ (left) and state $|2\rangle$ (right), and of
an equal mixture of the two states (middle).\label{f:imaging}}
\end{center}
\end{figure}

A degenerate Fermi gas of spin-polarized $^6$Li atoms was
prepared in an optical trap after laser cooling and sympathetic
cooling with sodium atoms~\cite{HGS03,ZAS05}. The population
imbalance $\delta $ of the two lowest hyperfine states
$|1\rangle$ and $|2\rangle$ was adjusted with a radio-frequency
sweep~\cite{ZSS06a}. Here, $\delta \equiv (N_1-N_2)/(N_1+N_2)$,
where $N_1$ and $N_2$ are the atom numbers in $|1\rangle$ and
$|2\rangle$, respectively. Interactions between these two states
were strongly enhanced near a broad Feshbach resonance at
$B_0=834$~G. The final evaporative cooling was performed at
$B=780$~G by lowering the trap depth. Subsequently, the
interaction strength was adiabatically changed to a target value
by adjusting the value of the magnetic-bias field $B$ with a ramp
speed of $\leq 0.4$~G/ms. For typical conditions, the total atom
number was $N_t=N_1+N_2\approx 1\times 10^7$ and the radial
(axial) trap frequency was $f_r=130$~Hz ($f_z=23$~Hz).

The condensate fraction in the imbalanced Fermi mixture was
determined via the rapid transfer technique~\cite{RGJ04,ZSS04}.
Immediately after turning off the trap, the magnetic field was
quickly ramped to $B=690$~G ($1/k_F a \approx 2.6$, where $k_F$
is defined as the Fermi momentum of a non-interacting equal
mixture with the same total atom number and $a$ is the scattering
length) in approximately $130~\mu$s. The density profile of the
expanding minority cloud was fit by a Gaussian for normal
components (thermal molecules and unpaired atoms) and a
Thomas-Fermi (TF) profile for the condensate~\cite{note1}.

The density difference between the two components was directly
measured using a phase-contrast imaging technique described in
Fig.~\ref{f:imaging}. In this imaging scheme, the signs of the
phase shifts due to the presence of atoms in each state are
opposite so that the resulting phase signal is proportional to the
density difference of the two states if the probe frequency is
adjusted properly~\cite{note2}. This technique allows us to
directly image the in situ distribution of the density difference
$n_d(\vec{r})$ between the two components and avoid the
shortcoming of previous studies~\cite{ZSS06a,PLK06,ZSS06b} where
two images were subtracted from each other.

\begin{figure}
\begin{center}
\includegraphics[width=3.3in]{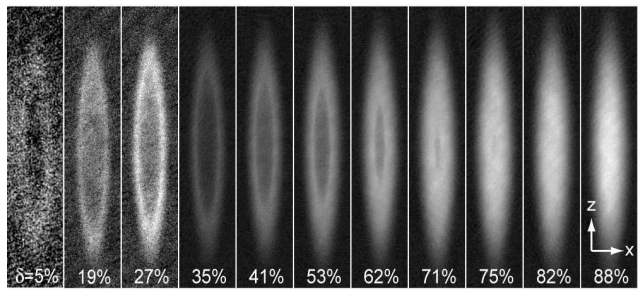}
\caption{In situ direct imaging of trapped Fermi gases with
population imbalance. The integrated 2D distributions of the
density difference $\tilde{n}_d(x,z)\equiv \int n_d(\vec{r})~dy$
were measured using phase-contrast imaging. The data were
obtained at $B=834~$G for total atom number $N_t\approx 1\times
10^7$ and various population imbalance $\delta$. For $\delta \leq
75\%$, a distinctive core was observed showing the shell
structure of the cloud. The field of view for each image is
$160~\mu$m$~\times~800~\mu$m. The three left-most images are
displayed with different contrast levels for clarity. The image
with $\delta=5(3)\%$ was taken for $N_t\approx 1.7 \times 10^7$.
\label{f:rawP}}
\end{center}
\end{figure}

For a partially superfluid imbalanced mixture, a shell structure
was observed in the in situ phase-contrast images
(Fig.~\ref{f:rawP}). Since the image shows the column density
difference (the 3D density difference integrated along the
$y$-direction of the imaging beam), the observed depletion in the
center indicates a 3D shell structure with even stronger
depletion in the central region. The size of this inner core
decreases for increasing imbalance and the core shows a
distinctive boundary until it disappears for large imbalance. We
observe this shell structure even for very small imbalances, down
to $5(3)\%$, which excludes a homogeneous superfluid state at
this low imbalance, contrary to the conclusions in
Ref.~\cite{PLK06}.

\begin{figure}
\begin{center}
\includegraphics[width=2.7in]{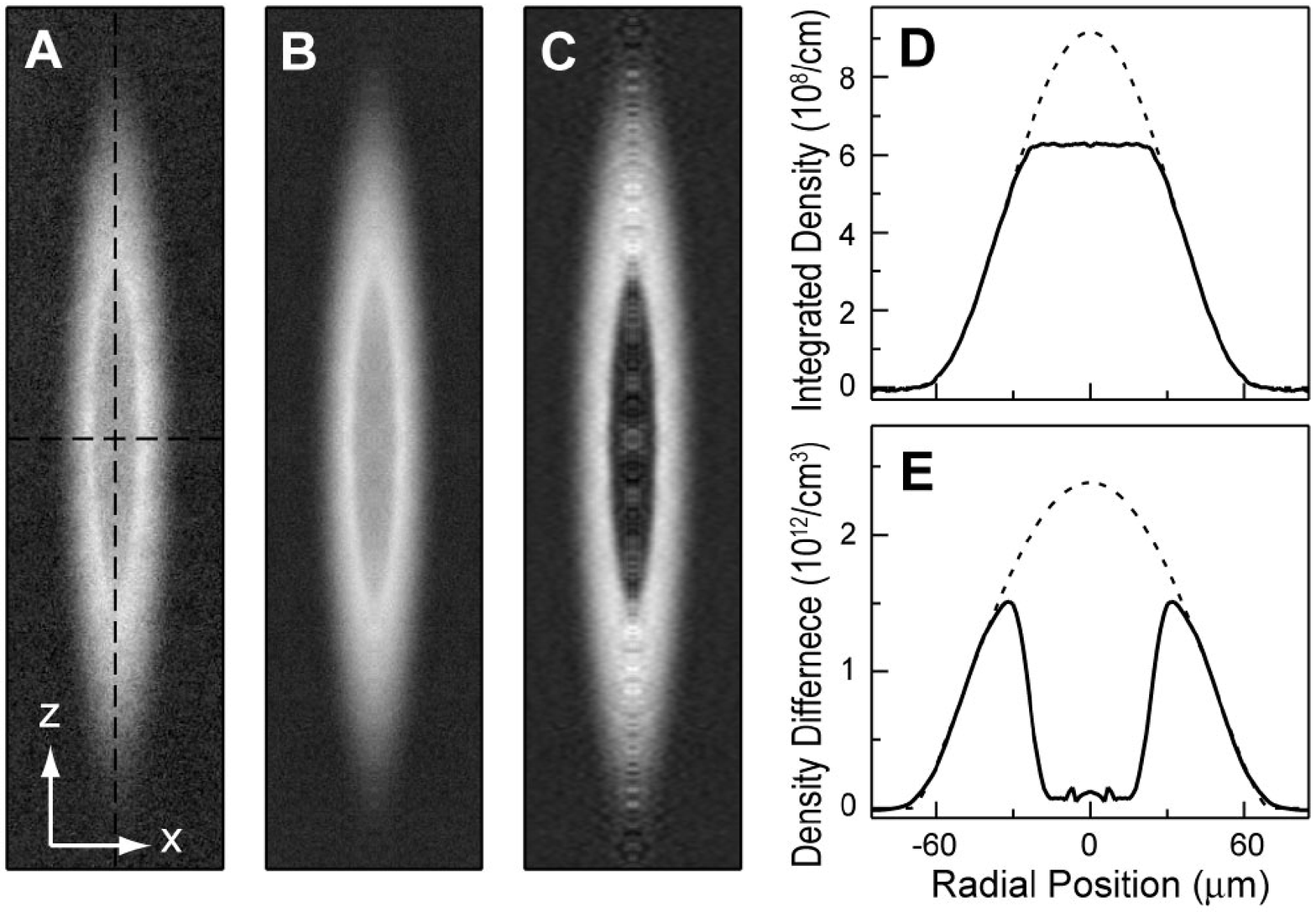}
\caption{Reconstruction of 3D distributions from their integrated
2D distributions. (a) An integrated 2D distribution $\tilde{n}_d$
with $\delta=58\%$ at $B=834~$G. (b) A less noisy distribution
was obtained by averaging four quadrants with respect to the
central dashed lines in (a). (d) The 1D profile obtained by
integrating the averaged distribution along $z$-direction shows
the `flat-top' feature. (c) The 2D-cut, $n_d(x,y=0,z)$, of the 3D
distribution $n_d(\vec{r})$ was reconstructed by applying the
inverse Abel transformation to the averaged distribution,
assuming only cylindrical symmetry along the $z$-direction. (e)
The radial profile of the central section of the reconstructed 3D
distribution in $xy$-plane. The dashed lines in (d) and (e) are
fits to the profiles' wings using a Thomas-Fermi (TF)
distribution.\label{f:abel}}
\end{center}
\end{figure}

The reconstructed 3D profile of the density difference shows that
the two components in the core region have equal densities. We
reconstruct 3D profiles from the 2D distributions
$\tilde{n}_d(x,z)$ of the column density difference using the
inverse Abel transformation (Fig.~\ref{f:abel})~\cite{Bra86}. The
only assumption employed in this process is that of cylindrical
symmetry of our trap along the axial $z$-direction. The two
transverse trap frequencies are equal to better than
$2\%$~\cite{ZAS05}. The reconstruction does not depend on the
validity of the local density approximation (LDA) or a harmonic
approximation for the trapping
potential~\cite{DSM06a,ZK06,IBL06}. A 1D profile obtained by
integrating $\tilde{n}_d$ along the axial $z$-direction
(Fig.~\ref{f:abel}(d)) shows a flat top distribution, which is
the expected outcome for a shell structure with an empty inner
region in a harmonic trap and assuming LDA.

\begin{figure}
\begin{center}
\includegraphics[width=2.7in]{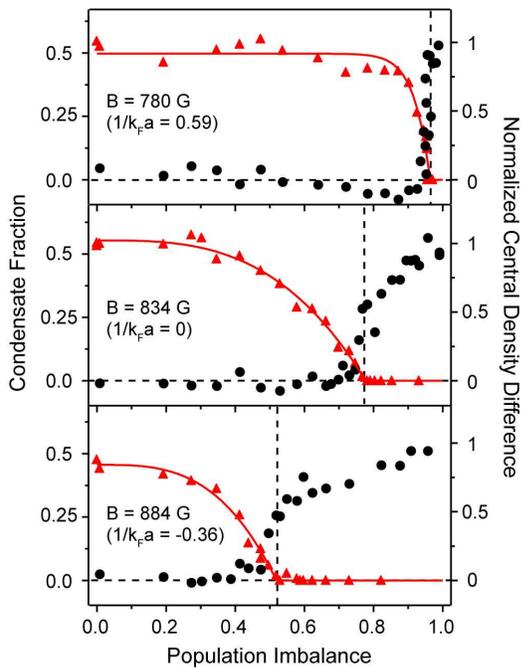}
\caption{(Color online) Phase separation in a strongly interacting
Fermi gas. Normalized central density difference $\eta$ (black
circle) and condensate fraction (red triangle) as functions of
imbalance $\delta$ for various interaction strengths. $\eta
\equiv n_{d0}/n_{0}$ where $n_{d0}$ was measured as the average
over the central region of $7~\mu$m $\times 40~\mu$m in a 2D-cut
of the 3D distribution (Fig.~\ref{f:abel}(c)) and $n_{0} = 3.3
\times 10^{12}$~cm$^{-3}$ is the calculated central density of a
fully polarized Fermi gas with $N_t = 1\times 10^7$. The
condensate fraction was determined from the bimodal distribution
of the minority component after a magnetic field sweep and
expansion. The solid line is a fit for the condensate fraction to
a threshold function $\propto (1-|\delta/\delta_c|^n)$. The
critical imbalance $\delta_c$ indicated by the vertical dashed
lines were 96\%, 77\%, and 51\% for $B=780$~G, $B=834$~G and
$B=884$~G, respectively.\label{f:phasep}}
\end{center}
\end{figure}

\begin{figure}
\begin{center}
\includegraphics[width=3.2in]{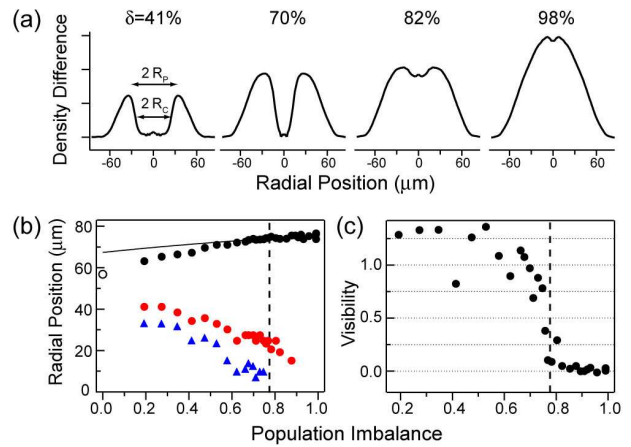}
\caption{(Color online) Characterization of the shell structure.
(a) Reconstructed 3D radial profiles at $B=834$~G. (b) Radius of
the majority component $R$ (black circle), peak position of the
density difference $R_p$ (red circle), and radius of the empty
core $R_c$ (blue triangle). $R$ was determined from the profiles'
wings using a fit to a zero-temperature TF distribution. The
solid line indicates the TF radius of the majority component in
an ideal non-interacting case. $R$ at $\delta=0$ (open circle)
was measured from an image taken with a probe frequency
preferentially tuned to one component. $R_c$ is defined as the
position of the half-peak value in the empty core region for
$\delta<\delta_c$. (c) Visibility of the core region is defined as
$\alpha \equiv (n_d(R_p)-n_{d0})/(n_d(R_p)+n_{d0})$.
\label{f:profiles}}
\end{center}
\end{figure}

\begin{figure}
\begin{center}
\includegraphics[width=3.2in]{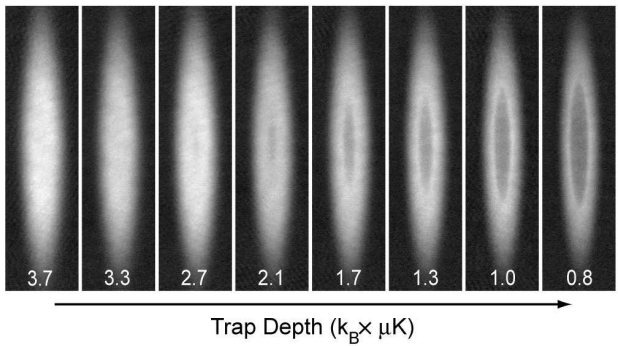}
\caption{Emergence of phase separation in an imbalanced Fermi
gas. The temperature of the cloud was controlled by varying the
final value of the trap depth $U_f$ in the evaporation process.
Phase-contrast images were taken after adiabatically ramping the
trap depth up to $k_B\times 3.7~\mu$K ($f_r=192$~Hz). The whole
evaporation and imaging process were performed at $B=834~$G ($N_t
\approx 1.7 \times 10^7$, $\delta \approx 56\%$). The field of
view for each image is $160~\mu$m$~\times~940~\mu$m. The vertical
and horizontal scale of the images differ by a factor of
1.5.\label{f:transition}}
\end{center}
\end{figure}

\begin{figure}
\begin{center}
\includegraphics[width=3in]{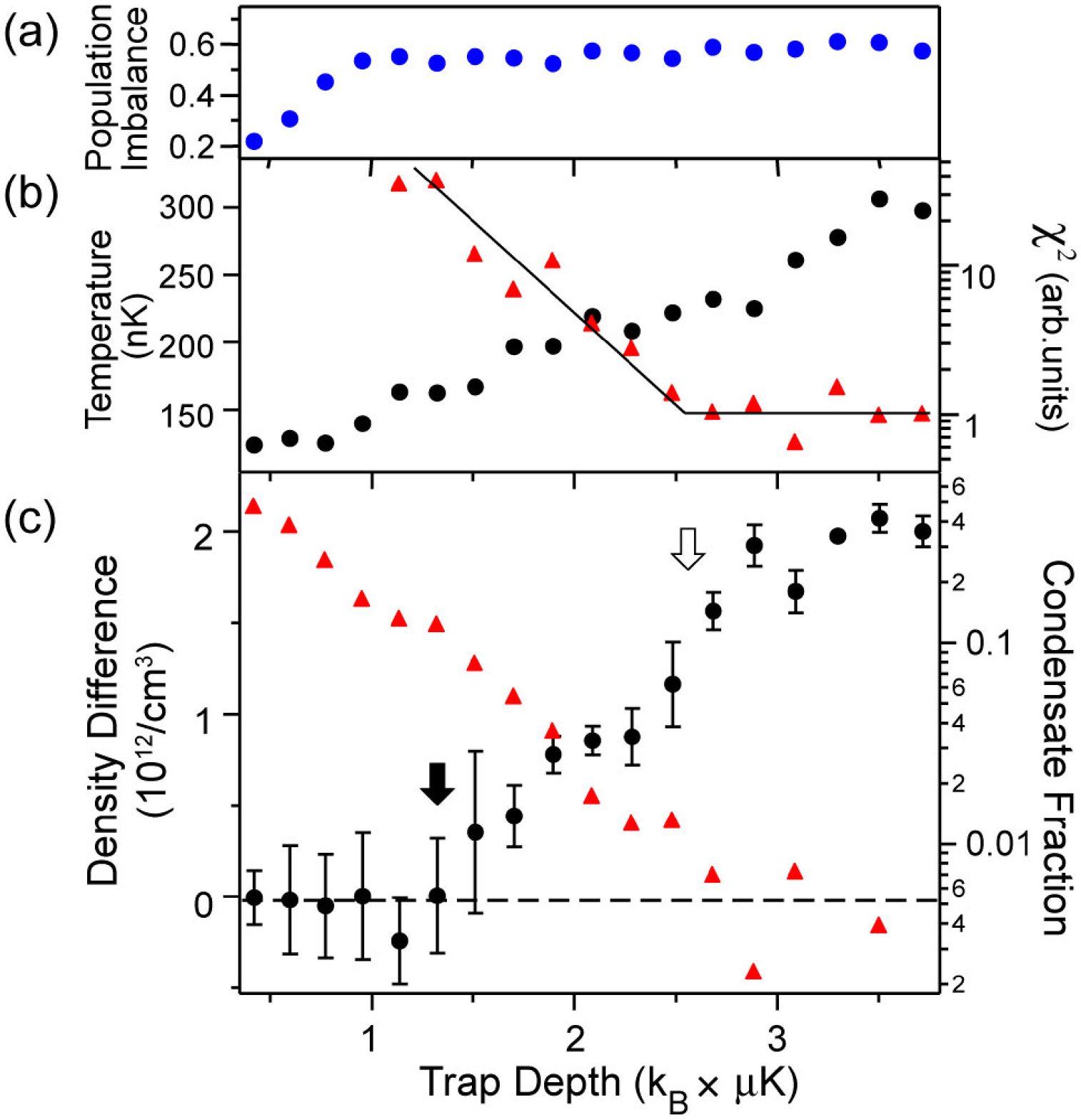}
\caption{(Color online) Phase transition in an imbalanced Fermi
gas. The phase transition shown in Fig.~\ref{f:transition} was
characterized by measuring (a) population imbalance $\delta$, (b)
temperature $T$ (black circle), $\chi^2$ for fitting the minority
cloud with a finite temperature Fermi-Dirac distribution (red
triangle), (c) central density difference $n_{d0}$ (black circle),
and condensate fraction (red triangle). The solid line is a guide
to the eye for $\chi^2$. When $U_f < k_B \times 1~\mu$K, $\delta$
decreased mainly due to loss of the majority component. The
number of the minority component was constant ($N_2 \approx
3.7\times 10^6$). $T$ was determined from the non-interacting
outer region of the majority cloud after 10~ms of ballistic
expansion~\cite{ZSS06b}. $T$ and $n_{d0}$ were averaged values of
three independent measurements. The open (solid) arrow in (c)
indicates the position for $T_c$ ($T^*$). See the text for the
definitions of $T_c$ and $T^*$.\label{f:temperature}}
\end{center}
\end{figure}

The presence of a core region with equal densities for the two
components was correlated with the presence of a pair condensate.
The density difference at the center $n_{d0}$ along with the
condensate fraction is shown as a function of the population
imbalance $\delta$ in Fig.~\ref{f:phasep}. As shown, there is a
critical imbalance $\delta_c$ where superfluidity breaks down due
to large imbalance~\cite{ZSS06a,ZSS06b}. In the superfluid
region, i.e., $\delta < \delta_c$, $n_{d0}$ vanishes, and for
$\delta
> \delta_c$, $n_{d0}$ rapidly increases with a sudden jump around $\delta \approx \delta_c$.
We observe a similar behavior throughout the strongly interacting
regime near the Feshbach resonance, $-0.4 < 1/k_Fa < 0.6$. This
observation clearly demonstrates that for this range of
interactions a paired superfluid is spatially separated from a
normal component of unequal densities.

The shell structure is characterized by the radius of the
majority component, the radial position $R_p$ of the peak in
$n_d(\vec{r})$, the size $R_c$ of the region where $n_d$ is
depleted, and the ``visibility" of the core region $\alpha \equiv
(n_d(R_p)-n_{d0})/(n_d(R_p)+n_{d0})$ (Fig.~\ref{f:profiles}). The
sudden drop of $\alpha$ around $\delta \approx \delta_c$ has the
same origin as the observed sudden jump of $n_{d0}$: the
breakdown of superfluidity (compare Fig.~\ref{f:phasep}). The
comparison between $R_p$ and $R_c$ shows that the boundary layer
between the superfluid and normal region is rather thin. It has
been suggested that the detailed shape of profiles in the
intermediate region could be used to identify exotic states such
as the Fulde-Ferrell-Larkin-Ovchinnikov (FFLO)
state~\cite{MMI05,YD06,MMI06}. This will be a subject of future
research.

The superfluid requires equal central densities in the strongly
interacting regime at our coldest temperatures. A normal
imbalanced Fermi mixture will have unequal densities. Thus, one
should expect that a visible change in the density difference
occurs as the temperature is lowered across the
normal-to-superfluid phase transition~\cite{BCR03,ZSS06b}. In
situ phase-contrast images of a cloud at various temperatures are
shown in Fig.~\ref{f:transition}. The temperature $T$ of the
cloud is controlled with the final value of the trap depth in the
evaporation process. The shell structure appears and becomes
prominent when $T$ decreases below a certain critical value. This
shell structure gives rise to the bimodal density profile of the
minority component that we observed after expansion from the trap
in our recent work~\cite{ZSS06b}. Here we show via \emph{in situ}
measurements that the onset of superfluidity is accompanied by a
pronounced change in the spatial density difference.

The phase transition is characterized in
Fig.~\ref{f:temperature}. As $T$ is lowered, $n_{d0}$ gradually
decreases from its plateau value and the condensate fraction
starts to increase. From the point of condensation (condensate
fraction $>1\%$) and deformation of the minority clouds ($\chi^2$
in Fig.~\ref{f:temperature}(b)), we determine the critical
temperature $T_c=0.13(2)~T_F$ for the imbalance of
$\delta=56(3)\%$. $T_F=1.7~\mu$K is the Fermi temperature of a
non-interacting equal mixture with the same total atom number.
The rise in $\chi^2$, the drop in $n_{d0}$ and the onset of
condensation are all observed at about the same temperature.
Better statistics are needed to address the question: whether some
weak expulsion of majority atoms from the center occurs already
slightly above $T_c$.

Below a certain temperature $T^*$, $n_{d0}$ reaches zero while
the condensate fraction is still increasing, implying that the
superfluid region of equal densities continues to expand spatially
with decreasing $T$. Full phase separation does not occur until
this temperature $T^*<T_c$ is reached. We interpret the state
between $T^*$ and $T_c$ as a superfluid of pairs coexisting with
polarized quasi-particle excitations. This is expected, since at
finite temperatures the BCS state of an equal mixture can
accommodate excess atoms as fermionic quasi-particle
excitations~\cite{YD06,CCH06a}. There is a finite energy given by
the pairing gap $\Delta(T)$ for those quasi-particles to exist in
the superfluid. The assumption that excess atoms should have
thermal energy $k_B T
> \Delta(T) $ to penetrate the superfluid region suggests the
relation between $T^*$ and $\Delta(T)$ to be $k_B T^* \approx
\Delta(T^*)$. From our experimental results, $T^*\approx
0.09~T_F$ and $\Delta(T^*) \approx h \times 3.3$~kHz.

In conclusion, we have observed phase separation of the
superfluid and the normal component in a strongly-interacting
imbalanced Fermi gas. The shell structure consisting of a
superfluid core of equal densities surrounded by a normal
component of unequal densities was clearly identified using in
situ phase-contrast imaging and 3D image reconstruction. The
phase-contrast imaging technique combined with the 3D
reconstruction process provides a new method to measure the in
situ density distribution, allowing direct comparison with
theoretical predictions.

We would like to thank G.~Campbell for critical reading of the
manuscript. This work was supported by the NSF, ONR, and NASA.


\begin{thebibliography}{24}
\bibitem{note1}
\bibinfo{journal}{This underestimates the ``true" condensate fraction since the bimodal
distribution used for fitting does not account for the repulsion
between the condensate and the normal components.}
\bibitem{note2}
\bibinfo{journal}{The probe frequency was detuned by
about 3 MHz from the exact center between the two resonance
frequencies in order to compensate the effect of absorption,
zeroing the optical signal in an equal mixture
(Fig.~\ref{f:imaging}).}
\end{thebibliography}
\end{document}